\documentclass[%
 aip,
 amsmath,amssymb,
 reprint,%
]{revtex4-2}
\usepackage{epsfig,amssymb,amsfonts}
\usepackage{amsmath,amsfonts,amssymb}
\usepackage{subfigure}
\usepackage{graphicx}
\usepackage{xcolor}
\input{epsf}
\usepackage{epstopdf}
\begin{document}

\title{Efficient four-wave mixing in four-subband semiconductor quantum wells using spatially modulated control fields with a linearly varying mixing angle}

\author{Dionisis Stefanatos}
\email{dionisis@post.harvard.edu}

\author{Foteini Avouri}

\author{Emmanuel Paspalakis}

\affiliation{Materials Science Department, School of Natural Sciences, University of Patras, Patras 26504, Greece}

\date{\today}

\begin{abstract}
In this article, we use spatially modulated control fields to increase the four-wave mixing efficiency
in a four-subband semiconductor asymmetric double quantum well, motivated by
similar works in atomic systems. Using a simplified version of the propagation equations, we show analytically that for control fields with constant amplitude and linearly varying mixing angle with the propagation distance, a conversion efficiency close to unity can
be achieved even for relatively short propagation distances. Subsequently, we confirm these results by numerically simulating the full set of propagation equations.
\end{abstract}

\maketitle

\section{Introduction}

Over the years, the phenomenon of quantum interference associated with Electromagnetically Induced Transparency (EIT) has driven considerable research efforts in quantum optics and quantum information science  \cite{Harris99,Paspalakis02,Lukin03,Fleischhauer05}, including some fascinating applications like the reduction of the light speed \cite{Hau99}. Among the various interesting applications, one of broad interest is the process of four-wave mixing (FWM). This procedure is encountered in a wide range of research fields, like storage and processing of quantum information \cite{Phillips11}, frequency conversion between light beams \cite{Lee16,Stefanatos20}, conversion between light beams carrying orbital angular momentum \cite{Hamedi19}, nonlinear optical amplification \cite{Li18}, etc.

By exploiting the analogy between atomic systems and electronic levels in semiconductor quantum wells, EIT has been demonstrated in such systems \cite{Serapiglia2000,Sadeghi1999,Sadeghi2000,Dynes2006,Frogley2006,Yang2009,Anton2008,Li2007,Hui2006,Kosionis2011,Paspalakis14}. Within this context, FWM emerging from quantum interference between the intersubband transitions in these systems has been investigated. In most of the corresponding studies \cite{Hao08,Yang09,She14,She15}, a four-subband configuration has been considered. In order to increase the relatively low FWM efficiency, in some works an extra (fifth) level is exploited \cite{Sun13,Liu15,Sun18}. In another study \cite{Liu14}, is taken advantage of the coupling of the energy levels to the continuum \cite{Faist97,Schmidt97,Faist99}. In the recent paper \cite{Yu19}, an extra coupling between the energy levels is utilized. The orbital angular momentum conversion between light beams propagating in semiconductor quantum wells has been proposed in the nice articles \cite{Zhang19,Wang20}. The common characteristic of all the above works is that, in order to create quantum interference and achieve FWM, they employ spatially independent pump (control) fields, which do not depend on the propagation distance.

In order to increase the conversion efficiency of the FWM process in semiconductor quantum well systems, here we propose to use spatially dependent control fields, which change with the propagation distance, motivated by analogous studies for atomic systems \cite{Lee16,Stefanatos20,Hamedi19}. Specifically, we consider the standard four-subband configuration and, as in Ref. \cite{Lee16}, the application of control fields with constant sum of intensities but mixing angle linearly varied with the propagation distance. Using an approximation of the propagation equations, we show analytically that FWM efficiency close to unity can be achieved even for relatively short propagation distances. Subsequently, we confirm these results by numerically simulating the full set of Maxwell-Schr\"{o}dinger equations.

The paper is organized as follows. In the next section we describe the four-subband semiconductor quantum well system and in Sec. \ref{sec:FWM} the FWM process using spatially dependent control fields. In Sec. \ref{sec:results} we present propagation simulation results using the complete set of Maxwell-Schr\"{o}dinger equations, while Sec. \ref{sec:conclusion} concludes the present work.

\section{Four-subband semiconductor asymmetric double quantum well}

\label{sec:4subband}

\begin{figure}[t]
\centering
\includegraphics[width=0.7\linewidth]{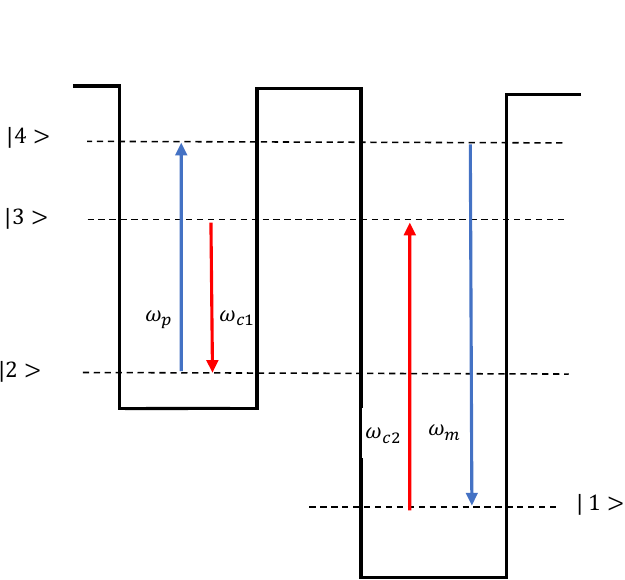}
\caption{Four-subband semiconductor asymmetric double quantum well. The strong control fields $\Omega_{c1}, \Omega_{c2}$ coherently prepare the medium where the weak probe pulse $\Omega_p$ is propagated and converted to the FWM pulse $\Omega_m$.}
\label{fig:ADQW}
\end{figure}

We consider an asymmetric semiconductor double quantum well with four subbands $|1\rangle, |2\rangle, |3\rangle, |4\rangle$, as shown in Fig. \ref{fig:ADQW}. For the energies of the four states we consider the values used in Ref. \cite{Liu14}, $E_1=51.53$ meV, $E_2=97.78$ meV, $E_3=191.3$ meV, and $E_4=233.23$ meV. Two strong continuous-wave (cw) control fields with central frequencies $\omega_{c1}, \omega_{c2}$ and wavevectors $k_{c1}, k_{c2}$ drive the transitions $|3\rangle\rightarrow|2\rangle$ and $|3\rangle\rightarrow|1\rangle$, respectively, coherently preparing the medium. A weak probe pulse with central frequency $\omega_p$ and wavevector $k_p$, which drives the $|2\rangle\rightarrow|4\rangle$ transition, is propagated through the coherently prepared medium. The weak four-wave mixing (FWM) pulse, with central frequency $\omega_m$ and wavevector $k_m$, arises from the $|4\rangle\rightarrow|1\rangle$ transition.

The interaction Hamiltonian under the rotating-wave and electric-dipole approximations is
\begin{eqnarray}
\label{Hamiltonian}
H/\hbar &=& \Delta_{2}|2\rangle\langle 2|+\Delta_{3}|3\rangle\langle 3|+\Delta_{4}|4\rangle\langle 4| \nonumber \\
  &-&\Big (\Omega_{c2}e^{ik_{c2}\cdot r}|3\rangle\langle 1|+\Omega_{c1}e^{ik_{c1}\cdot r}|3\rangle\langle 2| \nonumber \\
  &+& \Omega_{p}e^{ik_{p}\cdot r}|4\rangle\langle 2|+\Omega_{m}e^{ik_m\cdot r}|4\rangle\langle 1|+\mbox{H.c.} \Big ),
\end{eqnarray}
where the detunings are $\Delta_{3}=(E_3-E_1)-\omega_{c2}$, $\Delta_{2}=(E_2-E_1)-(\omega_{c2}-\omega_{c1})$, $\Delta_{4}=(E_4-E_1)-(\omega_{c2}-\omega_{c1}+\omega_{p})$, while the Rabi frequencies are expressed in terms of the slowly varying envelopes of the corresponding electric fields and transition dipole moments as $\Omega_p=\mu_{42}E_p/2\hbar$, $\Omega_{c1}=\mu_{32}E_{c1}/2\hbar$, $\Omega_{c2}=\mu_{31}E_{c2}/2\hbar$, $\Omega_{m}=\mu_{41}E_{m}/2\hbar$.
The state of the system can be expressed as
\begin{eqnarray}
\label{psi}
|\psi\rangle &=& A_1|1\rangle+A_2e^{i(k_{c2}-k_{c1})\cdot r}|2\rangle+A_3e^{ik_{c2}\cdot r}|3\rangle \nonumber \\
             &+& A_4e^{i(k_p-k_{c1}+k_{c2})\cdot r}|4\rangle,
\end{eqnarray}
where $A_i$, $i=1, 2, 3, 4$, are the time-dependent probability amplitudes of states $|i\rangle$. Using this expression in the Schr\"{o}dinger equation $i\hbar\partial|\psi\rangle/\partial t=H|\psi\rangle$
we end up with the following system for $A_i$,
\begin{subequations}
\label{amplitudes}
\begin{eqnarray}
i\frac{\partial A_1}{\partial t} & = & -\Omega^*_{c2}A_3-\Omega^*_me^{i\delta k\cdot r}A_4,\\
i\frac{\partial A_2}{\partial t} & = & \Delta_{2}A_2-i\gamma_2A_2-\Omega^*_{c1}A_3-\Omega^*_{p}A_4,\\
i\frac{\partial A_3}{\partial t} & = & \Delta_{3}A_3-i\gamma_3A_3-\Omega_{c2}A_1-\Omega_{c1}A_2,\\
i\frac{\partial A_4}{\partial t} & = & \Delta_{4}A_4-i\gamma_4A_4-\Omega_{p}A_2-\Omega_{m}e^{-i\delta k\cdot r}A_1.
\end{eqnarray}
\end{subequations}

Note that $\delta k=k_p-k_{c1}+k_{c2}-k_m$ expresses the phase mismatch, which subsequently is set equal to zero for simplicity, while we have manually inserted the decay rates $\gamma_i$, $i=2, 3, 4$. These rates can be expressed as a sum of two terms, $\gamma_i=\gamma_{il}+\gamma_{id}$, the fist term $\gamma_{il}$ expressing population decay because of longitudinal optical phonon emission and the second term $\gamma_{id}$
expressing dephasing due to acoustic phonon scattering. For these rates we will use the values from Ref. \cite{Liu14}. As discussed there, for the considered double quantum well in the absence of electronic continuum, it is $\gamma_{3l}\approx\gamma_{4l}=1$ meV, while for temperatures up to 10 K, where the electric density can be retained as low as $10^{24}$ m$^{-3}$, it is $\gamma_{3d}=0.32$ meV and $\gamma_{4d}=0.3$ meV. In total, $\gamma_3=\gamma_{3l}+\gamma_{3d}=1.32$ meV and $\gamma_4=\gamma_{4l}+\gamma_{4d}=1.3$ meV. On the other hand, $\gamma_2=\gamma_{2l}=2.36\times 10^{-6}$ $\mu$eV. From these values we conclude that the rates $\gamma_3, \gamma_4$ are close to each other while $\gamma_2\ll\gamma_3, \gamma_4$. This observation will be used in the next section in order to simplify the equations and obtain the spatially varying control fields.

We consider that the probe pulse $\Omega_p$ and the FWM pulse $\Omega_m$ are travelling in the $z$-direction, obeying in the slowly varying envelope approximation the wave equations
\begin{subequations}
\label{maxwell}
\begin{eqnarray}
\frac{\partial\Omega_p}{\partial z}+\frac{1}{c}\frac{\partial\Omega_p}{\partial t}&=&i\kappa_pA_4A^*_2,\\
\frac{\partial\Omega_m}{\partial z}+\frac{1}{c}\frac{\partial\Omega_m}{\partial t}&=&i\kappa_mA_4A^*_1,
\end{eqnarray}
\end{subequations}
where the propagation constants are $\kappa_p=N\omega_p|\mu_{42}|^2/2\hbar\varepsilon_0 c$ and $\kappa_m=N\omega_m|\mu_{41}|^2/2\hbar\varepsilon_0 c$, $N$ being the electron concentration in the semiconductor quantum wells. We will use the value $\kappa_p=\kappa_m=\kappa=9.6\times 10^3$ meV/$\mu$m from Ref. \cite{Liu14}. If we define the density matrix through the relations
\begin{equation}
\label{dm_element}
\rho_{ij}=A_iA^*_j,\quad i, j= 1, 2, 3, 4,
\end{equation}
then the above propagation equations become
\begin{subequations}
\label{maxwell1}
\begin{eqnarray}
\frac{\partial\Omega_p}{\partial z}+\frac{1}{c}\frac{\partial\Omega_p}{\partial t}&=&i\kappa\rho_{42},\\
\frac{\partial\Omega_m}{\partial z}+\frac{1}{c}\frac{\partial\Omega_m}{\partial t}&=&i\kappa\rho_{41}.
\end{eqnarray}
\end{subequations}

\section{Four-wave mixing using spatially modulated fields}

\label{sec:FWM}

In this section we make some simplifying assumptions to derive spatially modulated control fields $\Omega_{c1}, \Omega_{c2}$ which accomplish efficient FWM. These fields are tested in the next section using the full set of quantum (\ref{amplitudes}) and Maxwell (\ref{maxwell}) equations, without the simplifications made for their derivation. We start by deriving equations for the time evolution of the matrix elements $\rho_{42}, \rho_{41}$, appearing in the right hand side of the propagation equations (\ref{maxwell1}). Using Eqs. (\ref{amplitudes}) under the phase matching condition $\delta k=0$ we find
\begin{subequations}
\label{rho}
\begin{eqnarray}
\frac{\partial \rho_{42}}{\partial t} & = & i\Omega_p\rho_{22}+i\Omega_m\rho_{12}-i\Omega_{c1}\rho_{43}-i\Omega_p\rho_{44}\nonumber\\
                                      & - & [\gamma_2+\gamma_4+i(\Delta_4-\Delta_2)]\rho_{42},\\
\frac{\partial \rho_{41}}{\partial t} & = & i\Omega_m\rho_{11}+i\Omega_p\rho_{21}-i\Omega_{c2}\rho_{43}-i\Omega_m\rho_{44}\nonumber\\
                                      & - & (\gamma_4+i\Delta_4)\rho_{41}.
\end{eqnarray}
\end{subequations}
Regarding the decay rates, the approximations dictated by the values given in the previous section are $\gamma_2=0$ and $\gamma_3=\gamma_4=\gamma=1.3$ meV. Additionally, we take zero detunings $\Delta_{2}=\Delta_{3}=\Delta_4=0$. Under these assumptions, the above equations become
\begin{subequations}
\label{rho1}
\begin{eqnarray}
\frac{\partial \rho_{42}}{\partial t} & = & i\Omega_p\rho_{22}+i\Omega_m\rho_{12}-i\Omega_{c1}\rho_{43}-i\Omega_p\rho_{44}-\gamma\rho_{42},\\
\frac{\partial \rho_{41}}{\partial t} & = & i\Omega_m\rho_{11}+i\Omega_p\rho_{21}-i\Omega_{c2}\rho_{43}-i\Omega_m\rho_{44}-\gamma\rho_{41}.
\end{eqnarray}
\end{subequations}

We consider that the Rabi frequencies of the probe and FWM pulses, $\Omega_p, \Omega_m$, are much smaller than that of the cw pump fields $\Omega_{c1}, \Omega_{c2}$. The strong control fields are applied to the ``lower" $\Pi$-subsystem formed by the levels $|1\rangle, |2\rangle, |3\rangle$ and prepare it in the coherent dark state
\begin{equation}
\label{dark}
|\psi_d\rangle = \cos\theta|2\rangle-\sin\theta|1\rangle,
\end{equation}
where $\theta(z)$ is the spatially dependent mixing angle of the control fields defined by the relations
\begin{equation}
\label{controls}
\Omega_{c1}(z)=\Omega\sin{\theta(z)},\quad\Omega_{c2}(z)=\Omega\cos{\theta(z)},
\end{equation}
and  $\Omega$ is their constant amplitude. Thus, the weak probe pulse $\Omega_p$, which interacts with the ``upper" $\Pi$-subsystem formed by the levels $|1\rangle, |2\rangle, |4\rangle$, propagates in a coherently prepared medium. Since the probe and FWM fields are weak compared to the control fields, $\Omega_p, \Omega_m\ll \Omega$, during the propagation the populations $\rho_{11}, \rho_{22}$ and the coherence $\rho_{21}$ change little compared to their values in the initial dark state (\ref{dark}), thus we have
\begin{equation}
\label{pop_coh}
\rho_{11}\approx\sin^2\theta,\quad\rho_{22}\approx\cos^2\theta,\quad\rho_{21}\approx -\sin\theta\cos\theta.
\end{equation}
Additionally, the higher levels are hardly excited, thus we can assume to first order that $\rho_{43}\approx 0, \rho_{44}\approx 0$.
Using the above approximations, if we solve Eqs. (\ref{rho1}) for the elements $\rho_{42}, \rho_{41}$ in steady state we get, to first order with respect to the probe and FWM fields,
\begin{equation}
\label{rho_omega}
\left(
\begin{array}{c}
  \rho_{42}\\
  \rho_{41}
\end{array}
\right)
=
\frac{i}{\gamma}
\left(
\begin{array}{cc}
  \cos^2{\theta} & -\sin{\theta}\cos{\theta}\\
  -\sin{\theta}\cos{\theta} & \sin^2{\theta}
\end{array}
\right)
\left(
\begin{array}{c}
  \Omega_p\\
  \Omega_m
\end{array}
\right).
\end{equation}

If we substitute the above expressions for $\rho_{42}, \rho_{41}$ in Eqs. (\ref{maxwell1}), then we end up in the steady state with a pair of coupled equations describing the propagation of probe and FWM fields
\begin{equation}
\label{omega}
\frac{\partial}{\partial z}
\left(
\begin{array}{c}
\Omega_p\\
  \Omega_m
\end{array}
\right)
=
-\frac{\kappa}{\gamma}
\left(
\begin{array}{cc}
  \cos^2{\theta} & -\sin{\theta}\cos{\theta}\\
  -\sin{\theta}\cos{\theta} & \sin^2{\theta}
\end{array}
\right)
\left(
\begin{array}{c}
  \Omega_p\\
  \Omega_m
\end{array}
\right).
\end{equation}
For the normalized propagation distance
\begin{equation}
\label{zeta}
\zeta=\frac{2\kappa}{\gamma}z
\end{equation}
the propagation equations become
\begin{equation}
\label{system}
\frac{\partial}{\partial \zeta}
\left(
\begin{array}{c}
  \Omega_p\\
  \Omega_m
\end{array}
\right)
=
-\frac{1}{2}
\left(
\begin{array}{cc}
  \cos^2{\theta} & -\sin{\theta}\cos{\theta}\\
  -\sin{\theta}\cos{\theta} & \sin^2{\theta}
\end{array}
\right)
\left(
\begin{array}{c}
  \Omega_p\\
  \Omega_m
\end{array}
\right),
\end{equation}
where the factor of two in Eq. (\ref{zeta}) is selected so the analysis of Ref. \cite{Stefanatos20} can be immediately applied. The propagation takes place from $\zeta=0$ ($z=0$) to $\zeta=\alpha$, where
\begin{equation}
\alpha=\frac{2\kappa }{\gamma}Z
\end{equation}
corresponds to the final distance $z=Z$.

For a spatially varying $\theta(\zeta)$, it is more suitable to use the adiabatic basis of the propagation matrix in Eq. (\ref{system}), as it will become apparent below. The eigenstates, which correspond to eigenvalues $0$ and $-1/2$, are
\begin{equation}
\label{eigenstates}
\psi_0=\left(
\begin{array}{c}
  \sin{\theta}\\
  \cos{\theta}
\end{array}
\right),
\quad
\psi_{-1/2}=\left(
\begin{array}{c}
  \cos{\theta}\\
  -\sin{\theta}
\end{array}
\right).
\end{equation}
The transformation to the adiabatic basis is
\begin{equation}
\label{transformation}
\left(
\begin{array}{c}
  y\\
  x
\end{array}
\right)
=
\left(
\begin{array}{cc}
  \sin{\theta} & \cos{\theta}\\
  \cos{\theta} & -\sin{\theta}
\end{array}
\right)
\left(
\begin{array}{c}
  \Omega_p\\
  \Omega_m
\end{array}
\right),
\end{equation}
while the inverse transformation is
\begin{equation}
\label{inverse}
\left(
\begin{array}{c}
  \Omega_p\\
  \Omega_m
\end{array}
\right)
=
\left(
\begin{array}{cc}
  \sin{\theta} & \cos{\theta}\\
  \cos{\theta} & -\sin{\theta}
\end{array}
\right)
\left(
\begin{array}{c}
   y\\
   x
\end{array}
\right).
\end{equation}
From Eqs. (\ref{system}), (\ref{transformation}) and (\ref{inverse}), we get
\begin{equation}
\label{adiabatic}
\left(
\begin{array}{c}
  \dot{y}\\
  \dot{x}
\end{array}
\right)
=
\left(
\begin{array}{cc}
  0 & -u\\
  u & -\frac{1}{2}
\end{array}
\right)
\left(
\begin{array}{c}
  y\\
  x
\end{array}
\right),
\end{equation}
with $u(\zeta)$ being the derivative of the mixing angle
\begin{equation}
\label{theta}
\dot{\theta}=-u.
\end{equation}

\begin{figure}[t]
 \centering
		\begin{tabular}{cc}
     	\subfigure[$\ $]{
	            \label{fig:eff1}
	            \includegraphics[width=.5\linewidth]{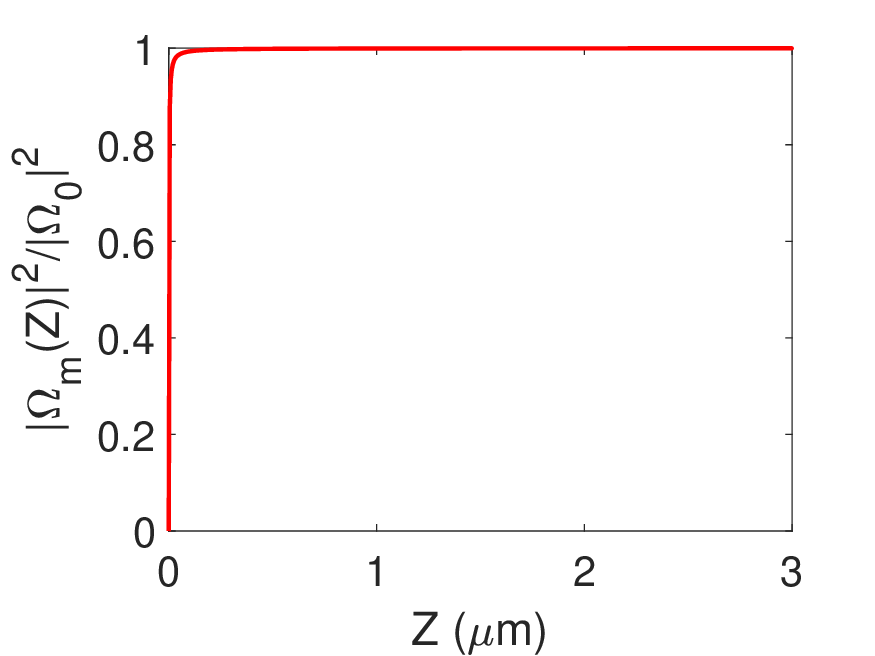}} &
        \subfigure[$\ $]{
	            \label{fig:eff2}
	            \includegraphics[width=.5\linewidth]{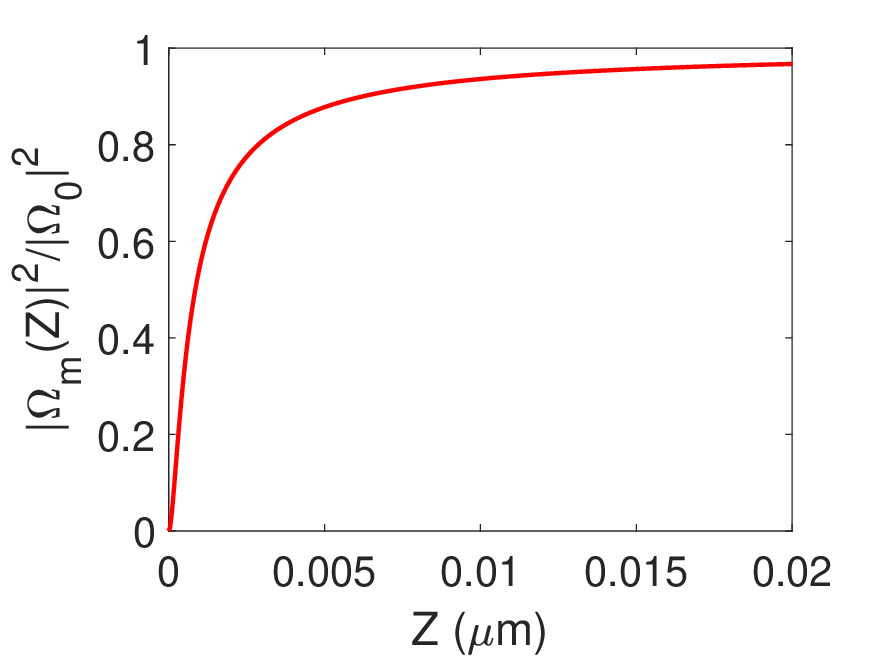}}
		\end{tabular}
\caption{(a) FWM efficiency for propagation distances up to $Z=3$ $\mu$m. (b) FWM efficiency for propagation distances up to $Z=20$ nm.}
\label{fig:efficiency}
\end{figure}

Now observe that if $\Omega_p(0)=\Omega_0, \Omega_m(0)=0$ and the boundary conditions of $\theta$ are selected as
\begin{equation}
\label{b_theta}
\theta(0)=\frac{\pi}{2},\quad\theta(\alpha)=0
\end{equation}
then from transformation (\ref{transformation}) we find
\begin{equation}
\label{initial}
x(0)=0,\quad y(0)=\Omega_p(0)=\Omega_0
\end{equation}
and
\begin{equation}
\label{final}
x(\alpha)=\Omega_p(\alpha),\quad y(\alpha)=\Omega_m(\alpha).
\end{equation}
If we consider a linear variation of $\theta$ \cite{Lee16,Stefanatos20}
\begin{equation}
\label{linear_theta}
\theta(\zeta)=\frac{\pi}{2}\left(1-\frac{\zeta}{\alpha}\right),
\end{equation}
then for large normalized propagation distance $\alpha\gg 1$ the mixing angle $\theta$ is slowly decreased form $\pi/2$ to $0$, with constant rate
\begin{equation}
u=|\dot{\theta}|=\frac{\pi}{2\alpha}=\frac{\pi\gamma}{4\kappa Z}.
\end{equation}
In this case the evolution is adiabatic and $y(\zeta)$ is approximately maintained constant, so $\Omega_m(\alpha)=y(\alpha)\approx y(0)=\Omega_p(0)=\Omega_0$. The adiabatic transformation of the probe field to the FWM field occurs through the eigenstate $\psi_0$.
Note that for constant $u$, no matter how large, the adiabatic system (\ref{adiabatic}) can be easily integrated and, using also Eqs. (\ref{initial}), (\ref{final}), we obtain the FWM efficiency for the normalized propagation distance $\zeta=\alpha$
\begin{equation}
\label{linear_eff}
\frac{|\Omega_m(\alpha)|^2}{|\Omega_0|^2}=e^{-\eta\alpha}\left[\cosh{(\rho\alpha)}+\frac{\eta}{2\rho}\sinh{(\rho\alpha)}\right]^2,
\end{equation}
with
\begin{equation*}
\eta=\frac{1}{2},\quad\rho=\sqrt{\left(\frac{\eta}{2}\right)^2-u^2}.
\end{equation*}
In the limits of short and long propagation distances we get the expressions
\begin{equation}
\label{limits_const}
\frac{|\Omega_m(\alpha)|^2}{|\Omega_0|^2}=\left\{\begin{array}{cl} \frac{1}{4\pi^2}\alpha^2, & \alpha\ll 1 \\1-\frac{\pi^2}{\alpha}, & \alpha\gg 1\end{array}\right.,
\end{equation}
from which it becomes obvious that the FWM efficiency tends to unity for large $\alpha$.
In Fig. \ref{fig:eff1} we plot the FWM efficiency (\ref{linear_eff}), obtained using the parameters of the system at hand, for propagation distances up to $Z=3$ $\mu$m. Observe that, even for very short distances, an efficiency close to unity is achieved. This is better seen in Fig. \ref{fig:eff2}, where is plotted the efficiency for propagation distances up to $Z=20$ nm. In the next section we test this efficiency, obtained using the simplified propagation equations, by simulating the full system of propagation equations.

\section{Simulation results and discussion}

\label{sec:results}

In this section we present simulation results of the propagation of probe and FWM pulses for various distances $Z$, using the full set of quantum and Maxwell equations (\ref{amplitudes}) and (\ref{maxwell}). We consider resonant fields, thus $\Delta_{2}=\Delta_{3}=\Delta_4=0$, decay rates $\gamma_2=2.36\times 10^{-6}$ $\mu$eV, $\gamma_3=1.32$ meV, $\gamma_4=1.3$ meV, and $\kappa_p=\kappa_m=9.6\times 10^3$ meV/$\mu$m, where note that all parameter values are taken from Ref. \cite{Liu14}. We consider spatially dependent control fields (\ref{controls}) with constant amplitude $\Omega=\gamma=1.3$ meV and linearly varying mixing angle with the propagation distance (\ref{linear_theta}). These fields prepare coherently the medium, and at its entrance $z=0$ the following weak Gaussian probe pulse is applied, while the FWM field is initially zero
\begin{subequations}
\label{pulses}
\begin{eqnarray}
\Omega_p(z=0, t) & = & \Omega_0e^{-\frac{(t-t_0)^2}{2\tau^2}},\label{zero_p} \\
\Omega_m(z=0, t) & = & 0, \label{zero_m}
\end{eqnarray}
where $\Omega_0=0.01\Omega=0.01\gamma$, $t_0=25\gamma^{-1}$ and $\tau=8\gamma^{-1}$.
\end{subequations}

\begin{figure*}[t]
 \centering
		\begin{tabular}{cc}
     	\subfigure[$\ $]{
	            \label{fig:c5}
	            \includegraphics[width=.5\linewidth]{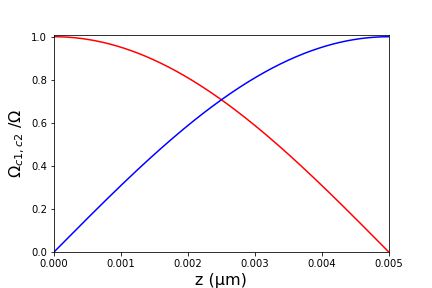}} &
        \subfigure[$\ $]{
	            \label{fig:m5}
	            \includegraphics[width=.5\linewidth]{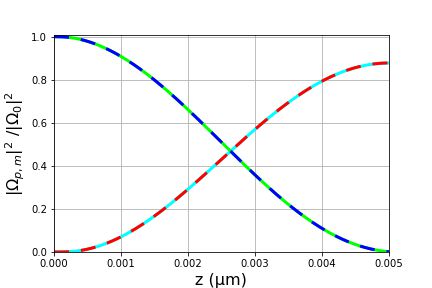}} \\
        \subfigure[$\ $]{
	            \label{fig:p5}
	            \includegraphics[width=.5\linewidth]{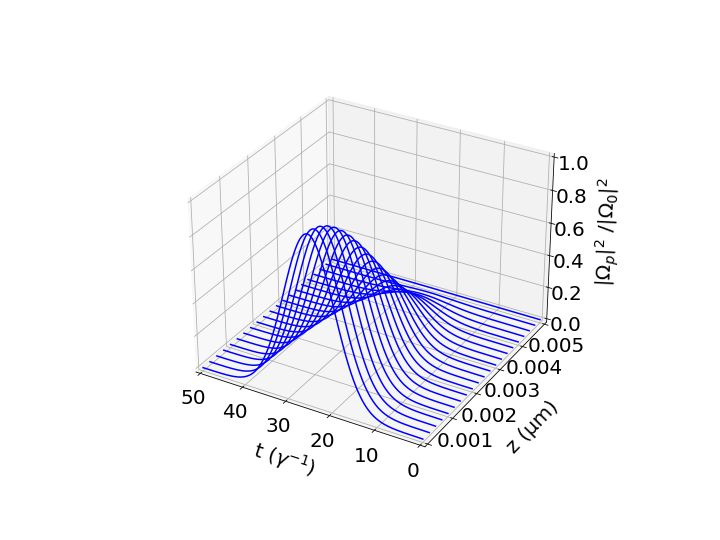}} &
        \subfigure[$\ $]{
	            \label{fig:mix5}
	            \includegraphics[width=.5\linewidth]{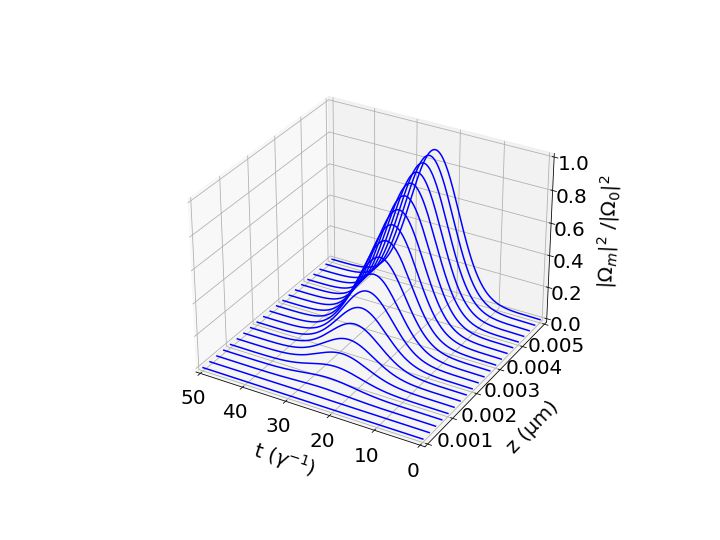}}
		\end{tabular}
\caption{Propagation for $Z = 5$ nm: (a) Spatial dependence of the cw control pulses $\Omega_{c1}$ (red), $\Omega_{c2}$ (blue). (b) Normalized peak intensity ($t=t_0$) for the probe pulse (approximate solid green, numerical dashed blue) and the FWM pulse (approximate solid cyan, numerical dashed red), along the propagation interval. (c) Propagation of the probe pulse (normalized intensity). (d) Propagation of the FWM pulse (normalized intensity).}
\label{fig:5nm}
\end{figure*}

\begin{figure*}[t]
 \centering
		\begin{tabular}{cc}
     	\subfigure[$\ $]{
	            \label{fig:c10}
	            \includegraphics[width=.5\linewidth]{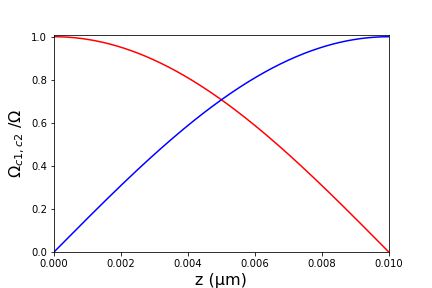}} &
        \subfigure[$\ $]{
	            \label{fig:m10}
	            \includegraphics[width=.5\linewidth]{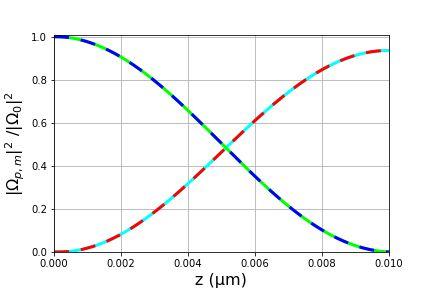}} \\
        \subfigure[$\ $]{
	            \label{fig:p10}
	            \includegraphics[width=.5\linewidth]{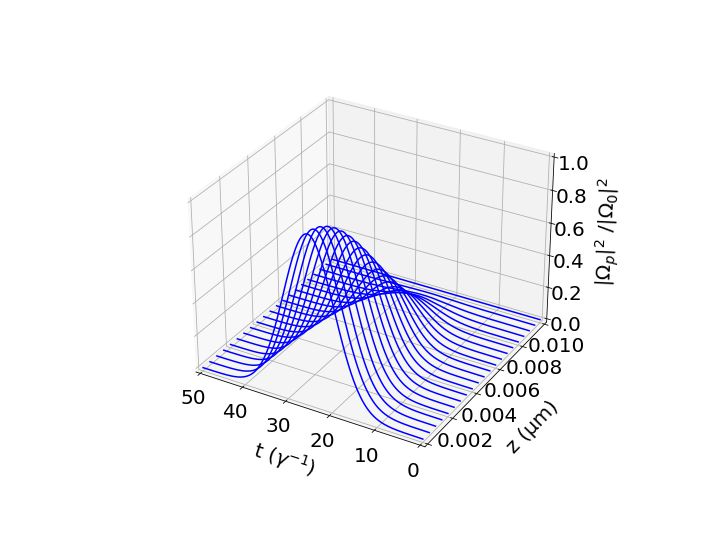}} &
        \subfigure[$\ $]{
	            \label{fig:mix10}
	            \includegraphics[width=.5\linewidth]{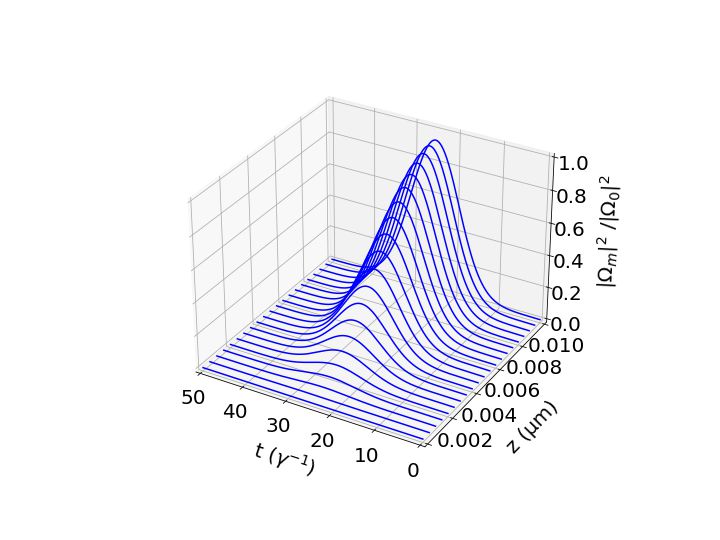}}
		\end{tabular}
\caption{Propagation for $Z = 10$ nm: (a) Spatial dependence of the cw control pulses $\Omega_{c1}$ (red), $\Omega_{c2}$ (blue). (b) Normalized peak intensity ($t=t_0$) for the probe pulse (approximate solid green, numerical dashed blue) and the FWM pulse (approximate solid cyan, numerical dashed red), along the propagation interval. (c) Propagation of the probe pulse (normalized intensity). (d) Propagation of the FWM pulse (normalized intensity).}
\label{fig:10nm}
\end{figure*}

\begin{figure*}[t]
 \centering
		\begin{tabular}{cc}
     	\subfigure[$\ $]{
	            \label{fig:c100}
	            \includegraphics[width=.5\linewidth]{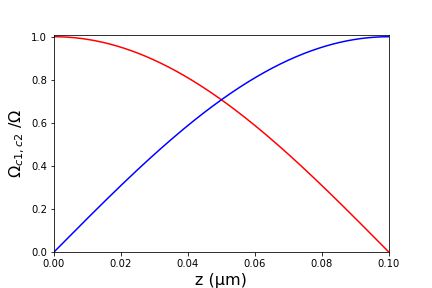}} &
        \subfigure[$\ $]{
	            \label{fig:m100}
	            \includegraphics[width=.5\linewidth]{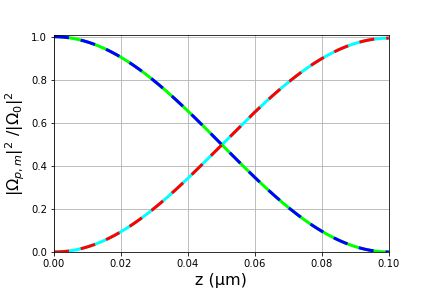}} \\
        \subfigure[$\ $]{
	            \label{fig:p100}
	            \includegraphics[width=.5\linewidth]{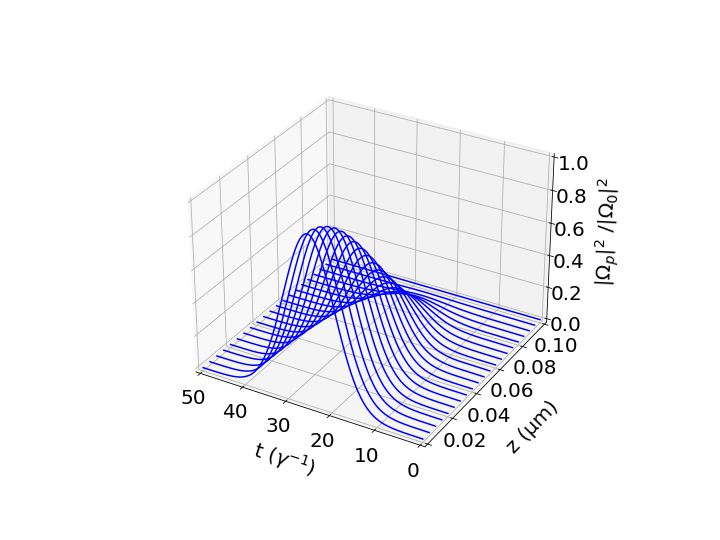}} &
        \subfigure[$\ $]{
	            \label{fig:mix100}
	            \includegraphics[width=.5\linewidth]{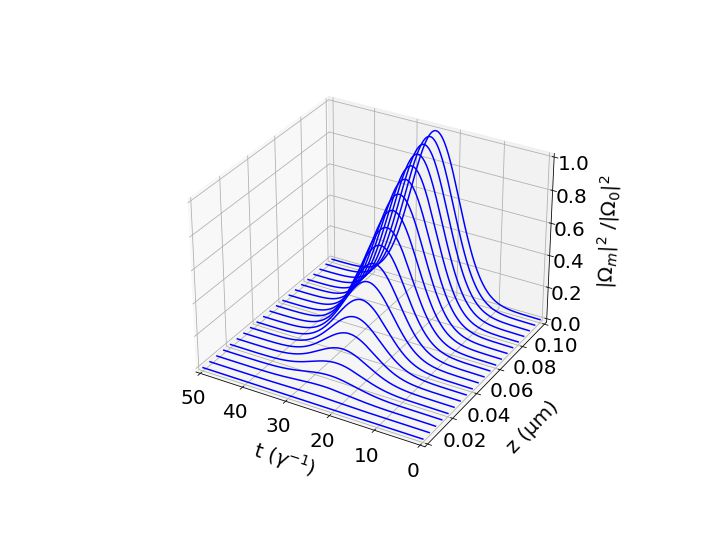}}
		\end{tabular}
\caption{Propagation for $Z = 0.1$ $\mu$m: (a) Spatial dependence of the cw control pulses $\Omega_{c1}$ (red), $\Omega_{c2}$ (blue). (b) Normalized peak intensity ($t=t_0$) for the probe pulse (approximate solid green, numerical dashed blue) and the FWM pulse (approximate solid cyan, numerical dashed red), along the propagation interval. (c) Propagation of the probe pulse (normalized intensity). (d) Propagation of the FWM pulse (normalized intensity).}
\label{fig:100nm}
\end{figure*}

\begin{figure*}[t]
 \centering
		\begin{tabular}{cc}
     	\subfigure[$\ $]{
	            \label{fig:c500}
	            \includegraphics[width=.5\linewidth]{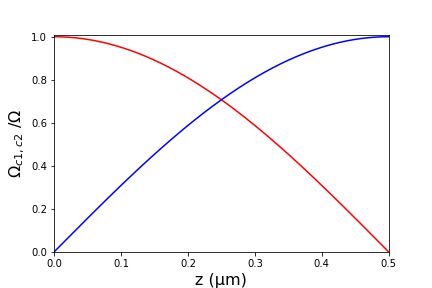}} &
        \subfigure[$\ $]{
	            \label{fig:m500}
	            \includegraphics[width=.5\linewidth]{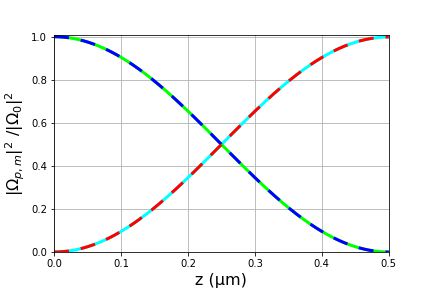}} \\
        \subfigure[$\ $]{
	            \label{fig:p500}
	            \includegraphics[width=.5\linewidth]{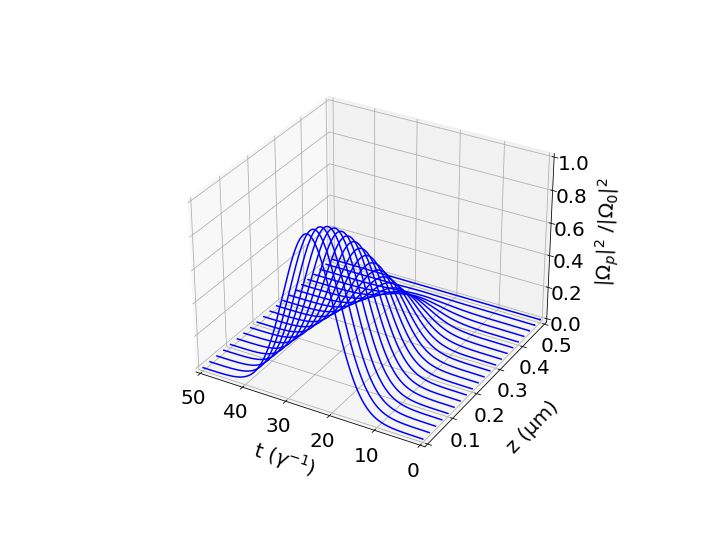}} &
        \subfigure[$\ $]{
	            \label{fig:mix500}
	            \includegraphics[width=.5\linewidth]{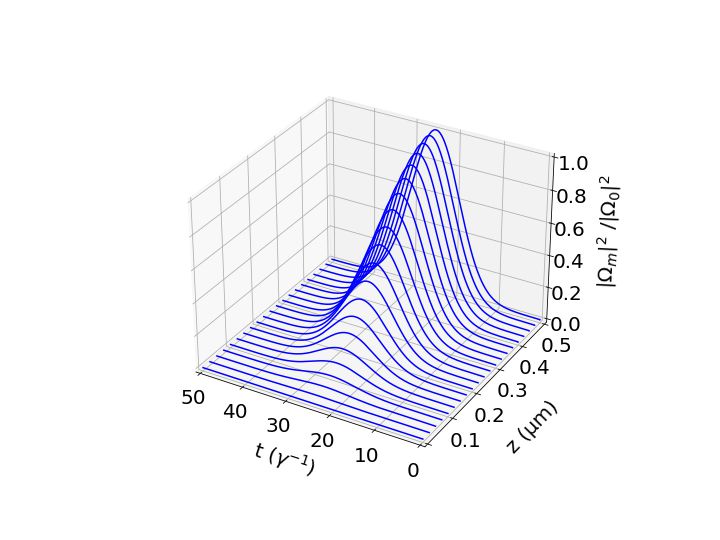}}
		\end{tabular}
\caption{Propagation for $Z = 0.5$ $\mu$m: (a) Spatial dependence of the cw control pulses $\Omega_{c1}$ (red), $\Omega_{c2}$ (blue). (b) Normalized peak intensity ($t=t_0$) for the probe pulse (approximate solid green, numerical dashed blue) and the FWM pulse (approximate solid cyan, numerical dashed red), along the propagation interval. (c) Propagation of the probe pulse (normalized intensity). (d) Propagation of the FWM pulse (normalized intensity).}
\label{fig:500nm}
\end{figure*}

\begin{figure*}[t]
 \centering
		\begin{tabular}{cc}
     	\subfigure[$\ $]{
	            \label{fig:c1000}
	            \includegraphics[width=.5\linewidth]{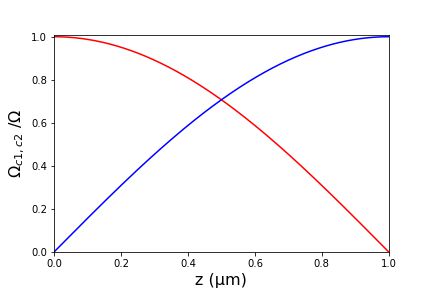}} &
        \subfigure[$\ $]{
	            \label{fig:m1000}
	            \includegraphics[width=.5\linewidth]{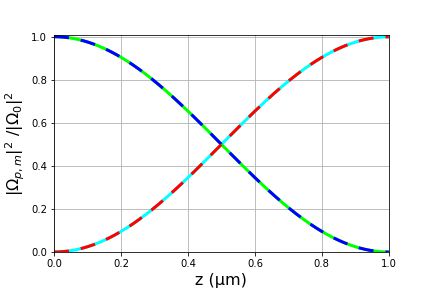}} \\
        \subfigure[$\ $]{
	            \label{fig:p1000}
	            \includegraphics[width=.5\linewidth]{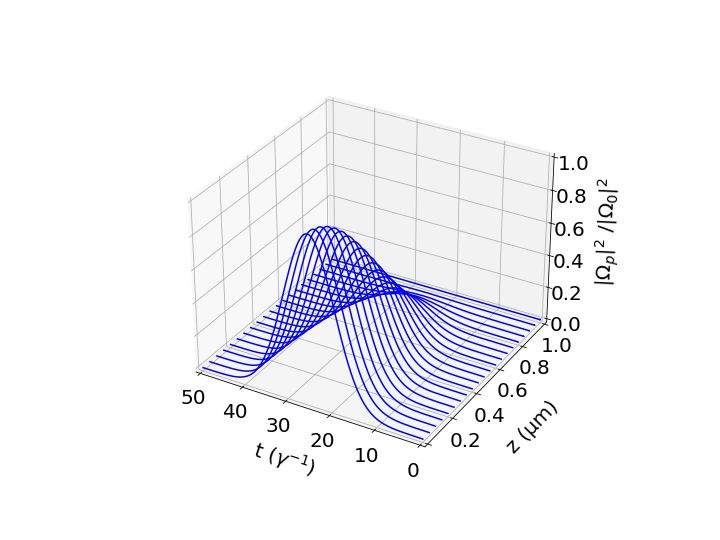}} &
        \subfigure[$\ $]{
	            \label{fig:mix1000}
	            \includegraphics[width=.5\linewidth]{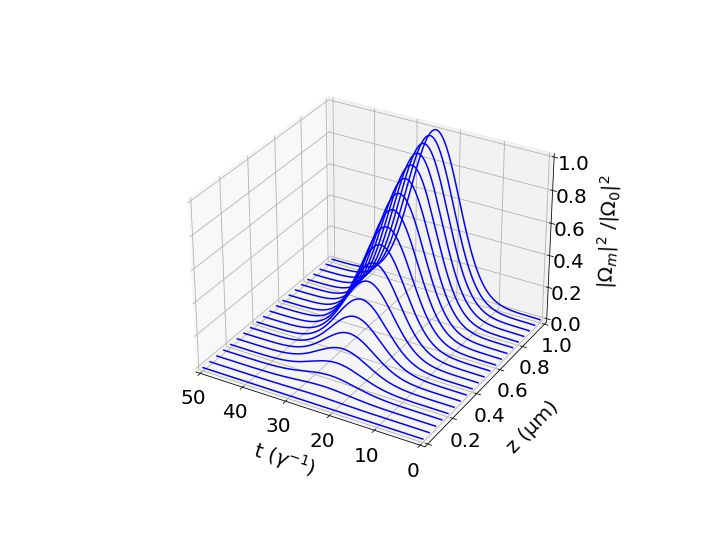}}
		\end{tabular}
\caption{Propagation for $Z = 1$ $\mu$m: (a) Spatial dependence of the cw control pulses $\Omega_{c1}$ (red), $\Omega_{c2}$ (blue). (b) Normalized peak intensity ($t=t_0$) for the probe pulse (approximate solid green, numerical dashed blue) and the FWM pulse (approximate solid cyan, numerical dashed red), along the propagation interval. (c) Propagation of the probe pulse (normalized intensity). (d) Propagation of the FWM pulse (normalized intensity).}
\label{fig:1000nm}
\end{figure*}

\begin{figure*}[t]
 \centering
		\begin{tabular}{cc}
     	\subfigure[$\ $]{
	            \label{fig:c3000}
	            \includegraphics[width=.5\linewidth]{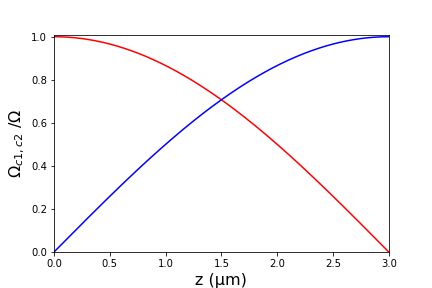}} &
        \subfigure[$\ $]{
	            \label{fig:m3000}
	            \includegraphics[width=.5\linewidth]{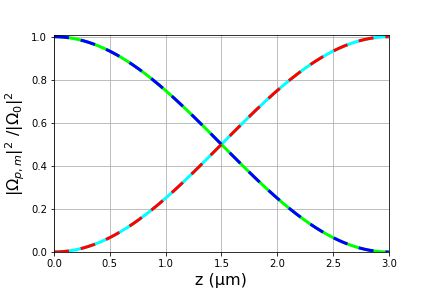}} \\
        \subfigure[$\ $]{
	            \label{fig:p3000}
	            \includegraphics[width=.5\linewidth]{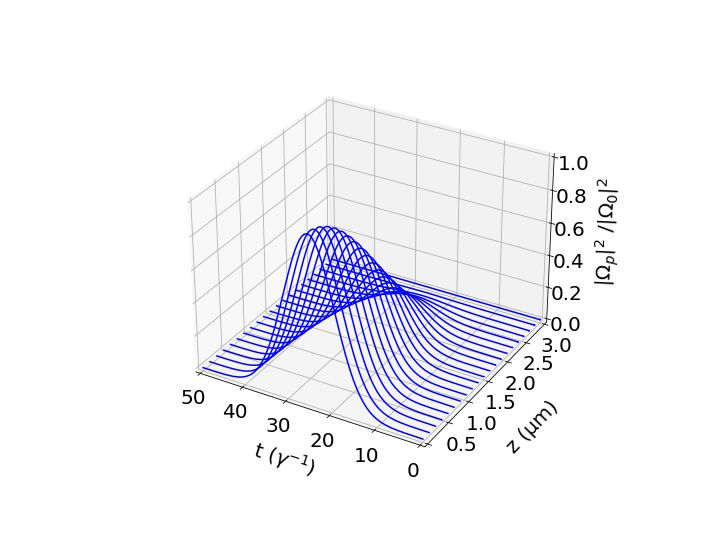}} &
        \subfigure[$\ $]{
	            \label{fig:mix3000}
	            \includegraphics[width=.5\linewidth]{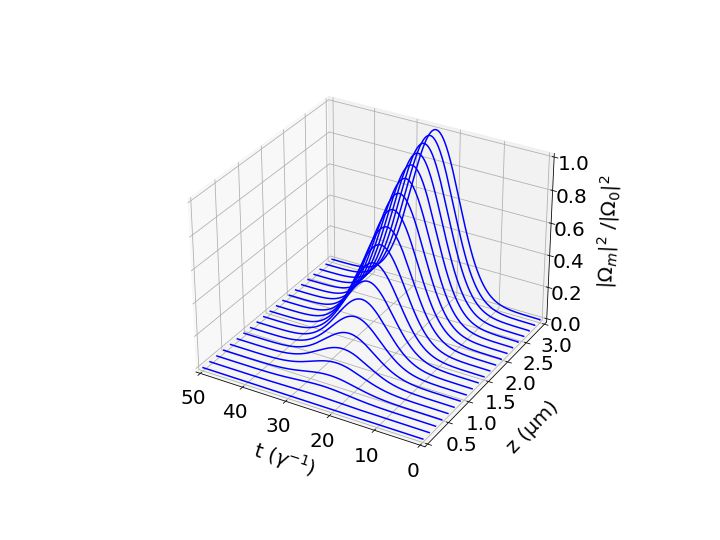}}
		\end{tabular}
\caption{Propagation for $Z = 3$ $\mu$m: (a) Spatial dependence of the cw control pulses $\Omega_{c1}$ (red), $\Omega_{c2}$ (blue). (b) Normalized peak intensity ($t=t_0$) for the probe pulse (approximate solid green, numerical dashed blue) and the FWM pulse (approximate solid cyan, numerical dashed red), along the propagation interval. (c) Propagation of the probe pulse (normalized intensity). (d) Propagation of the FWM pulse (normalized intensity).}
\label{fig:3000nm}
\end{figure*}

In Figs. \ref{fig:5nm}, \ref{fig:10nm}, \ref{fig:100nm}, \ref{fig:500nm}, \ref{fig:1000nm}, \ref{fig:3000nm}, we present simulation results for propagation distances $5$ nm, $10$ nm, $0.1$ $\mu$m, $0.5$ $\mu$m, $1$ $\mu$m, $3$ $\mu$m, respectively. For the shortest propagation distance $Z=5$ nm, in Fig. \ref{fig:c5} we display the spatially modulated control fields. In Fig. \ref{fig:m5} we show the normalized peak intensity at $t=t_0$ for the probe pulse (approximate solid green, numerical dashed blue) and the FWM pulse (approximate solid cyan, numerical dashed red), along the propagation interval. Observe the excellent agreement between the approximate efficiency obtained from the simplified model of the previous section and the numerical efficiency obtained from the simulation. In Figs. \ref{fig:p5} and \ref{fig:mix5} we plot the propagation of the normalized intensity for the probe and FWM pulses, respectively. The FWM efficiency at the final distance $Z=5$ nm is 0.8779. For the longer propagation distances displayed in the rest of the figures we also observe an excellent agreement between theory and simulation. The FWM efficiency obtained increases approaching unity: 0.9362 for $Z=10$ nm, 0.9933 for $Z=0.1$ $\mu$m, 0.9987 for $Z=0.5$ $\mu$m, 0.9993 for $Z=1$ $\mu$m, and 0.9998 for $Z=3$ $\mu$m. Note that the longer propagation distances may be more relevant for possible experimental realization of the proposed scheme, since it might be easier to implement the desired spatial modulation of the control fields over longer distances. A relevant experimental setup for the proposed scheme is that described in Refs. \cite{Wang20,Faist94}, where constant control fields are employed. The implementation of spatially varying control fields might seem challenging, but note that such fields have been experimentally realized in Ref. \cite{Lee16}. Even if the spatial profiles (\ref{controls}) of the control fields cannot be exactly implemented, a large conversion efficiency can still be obtained by using feasible Gaussian profiles mimicking STIRAP \cite{Lee16}.

\section{Summary}

\label{sec:conclusion}

In this work, we showed theoretically and with numerical simulations that, spatially dependent control fields which are properly modulated can be exploited to increase the four-wave mixing efficiency
in a four-subband semiconductor asymmetric double quantum well. This study was motivated by
similar works in atomic systems.

\section*{Acknowledgements}
The work of D.S. was funded by an Empirikion Foundation research grant.

\section*{DATA AVAILABILITY}

The data that support the findings of this study are available within the article.

\bibliographystyle{apsrev}
\bibliography{bibliography}

\end{document}